\documentclass[review]{article}
\usepackage[dvipdfm]{graphicx}
\usepackage{amsthm}
\usepackage[usenames]{color}
\usepackage{xspace,colortbl}
\usepackage{latexsym,url,amscd,amssymb}
\usepackage{amsmath,amsfonts}\numberwithin{equation}{section}
\usepackage{graphics}
\usepackage{epstopdf}
\usepackage{hyperref}
\usepackage{cases}
\usepackage{caption}
\usepackage{algorithm}
\usepackage{algpseudocode}
\usepackage{listings}
\usepackage{rotating}
\usepackage{framed}
\usepackage{booktabs}
\usepackage{indentfirst}
\usepackage{bm}
\usepackage[numbers,sort,compress]{natbib}
\usepackage{float}
\usepackage{subfig}
\usepackage{array}
\usepackage{multirow}
\usepackage{natbib}
\usepackage{ulem}
\usepackage{authblk}

\DeclareMathOperator*{\argmin}{arg\,min}

\def\prox{{\rm Prox}}

\def\B{{\mathcal B}}

\def\S{{\mathcal S}}

\newtheorem{theorem}{Theorem}[section]

\newtheorem{remark}{Remark}[section]

\date{} 
\author[a]{Mengjiao Shi}
\author[a,b]{Yunhai Xiao \thanks{Corresponding author: yhxiao@henu.edu.cn}}%
\affil[a]{School of Mathematics and Statistics, Henan University, Kaifeng 475000, China}
\affil[b]{Center for Applied Mathematics of Henan Province, Henan University, Zhengzhou 450046, China}

\title{An Efficient Dual ADMM for Huber Regression with Fused Lasso Penalty}

\begin{document} 
	\maketitle
	
	\begin{abstract}
		{
			The ordinary least squares estimate in linear regression is sensitive to the influence of errors with large variance, which reduces its robustness, especially when dealing with heavy-tailed errors or outliers frequently encountered in real-world scenarios. 
			To address this issue and accommodate the sparsity of coefficients along with their sequential disparities, we combine the adaptive robust Huber loss function with a fused lasso penalty. This combination yields a robust estimator capable of simultaneously achieving estimation and variable selection.
			Furthermore, we utilize an efficient alternating direction method of multipliers to solve this regression model from a dual perspective.
			The effectiveness and efficiency of our proposed approach is demonstrated through numerical experiments carried out on both simulated and real datasets.
		}
		
	\end{abstract}
	

	\section{Introduction}
	
	Consider the independent and identically distributed (i.i.d.) observations $\{(y_i, X_i)\}_{i=1}^n$ from $( y, X)$ which conform to the linear model
	$$ y=X\beta^*+\epsilon,$$ 
	where $ X=(X_1,X_2,\ldots,X_n)^{\top}\in \mathbb R^{n\times p}$ is  a predictor with $X_i\in\mathbb{R}^p$, $\beta^*=(\beta^*_1,\beta^*_2,\dots,\beta^*_p)^{\top}\in \mathbb R^p$ is a vector of regression coefficient, and $\epsilon\in\mathbb{R}^p$ is an error (noise) term satisfying  $\mathbb{E}(\epsilon| X)=0$. 
	Among the many statistical learning techniques available for estimating coefficients $ \beta^*$, the ordinary least squares method (OLS) is notable for its extensive use, attributed to its simplicity and effectiveness in model fitting. 
	However, the effectiveness of OLS, particularly in high-dimensional data settings, depends heavily on assuming errors that are light-tailed or even sub-Gaussian \cite{wainwright2019high}.  
	In fact, the sub-Gaussian tail assumption is often unrealistic in many practical applications as modern data are often corrupted by heavy-tailed noise and are collected under conditions of varying quality. 
	For instance, functional magnetic resonance imaging data often deviate from the expected Gaussian distribution. Wang et al. \cite{wangahigh} noted that certain gene expression levels display kurtosis values significantly exceeding three.   
	Heavy-tailed distributions are a primary characteristic of high-dimensional data, commonly found in modern statistical and machine learning problems. 
	Given the usual poor performance of ordinary least squares estimation  in such contexts, the development of robust statistical methods with strong theoretical properties has become increasingly important.

	To address this issue, a variety of robust regularization techniques have been extensively studied. Effective methods include quantile regression \cite{Koenker1978RegressionQ}, least absolute deviation (LAD) regression \cite{Bassett1978AsymptoticTO}, and Huber-type estimators. 
	Wang \cite{wang2013l1} studied $\ell_1$-penalized LAD regression and showed that the estimator achieves near oracle risk performance in high dimensional settings. Unfortunately, the LAD loss is not suitable for small errors, especially when the errors don't have heavy tail and are not affected by outliers.  
	In such cases, this estimator is expected to be less efficient than the OLS.  
	To adapt to different  error distributions and to enhance robustness in estimation, the Huber loss \cite{Huber1964RobustEO} has attracted considerable attention.
	It combines squared loss for smaller errors with absolute loss for larger errors, that is
	\begin{equation}
		\label{huber}
		h_{\tau}(x)=\left\{
		\begin{aligned}
			&\frac{1}{2}x^2, &&{\text{if}~|x|\leq\tau},\\
			&\tau\mid x\mid - \frac{1}{2}\tau^2, &&{\text{if}~|x| >\tau},
		\end{aligned}
		\right.
	\end{equation}
	where the degree of hybridization is controlled by parameter $\tau~ (\tau>0)$. To estimate the mean regression functions in ultrahigh dimensional settings, Fan et al. \cite{fan2017Estimation} proposed a penalized Huber loss with diverging parameter to reduce bias.  Additionally,  Fan et al.  \cite{sun2020adaptive} also used the Huber regression with an adaptive robustification parameter from a non-asymptotic perspective. 
	
	Variable selection is crucial in linear regression analysis, and employing a penalty term is essential for selecting variables, especially when the true underlying model exhibits sparsity.
	Tibshirani \cite{Tibshirani1996RegressionSA} introduced the lasso (least absolute shrinkage and selection operator) penalty, which is a regularization technique applying $\ell_1$-norm to shrink coefficients in least squares regression.
	Moreover, Tibshirani \cite{tibshirani2005sparsity} introduced a novel fused lasso  to penalize the $\ell_1$-norm of both coefficients and their successive differences, particularly suited for data where features can be meaningfully ordered.
	Since then, the fused lasso has garnered significant research attention, as evidenced by studies such as \cite{petersen2016fused, mao2021robust, corsaro2021fused,cui2021fused,degras2021sparse,li2019double}.
	Recently, combined with the fused lasso, Xin et al. \cite{xin2022robust} proposed an adaptive Huber regression, that is
	\begin{equation}\label{admod}
		\min\limits_{ \beta \in \mathbb{R}^p} 
		{\cal H}_{\tau}( \beta) + \lambda_1\| \beta\|_1 + \lambda_2\sum_{j=2}^{p}|\beta_{j} - \beta_{j-1}|,
	\end{equation}
	where $\lambda_1>0$ and $\lambda_2>0$ are tuning parameters and  ${\cal H}_{\tau}( \beta)$	is an empirical Huber loss function for $\beta$. In fact, if the third term is vanished, it corresponds to the adaptive Huber regression of Fan et al. \cite{sun2020adaptive}. 
	It is important to emphasize that the method described by \eqref{admod} is adaptive to data contaminated by both light-tailed and heavy-tailed distribution noise. However, the model \eqref{admod} is convex yet non-smooth, which presents substantial challenges implementation.
	
	It should be noted that there are numerous practical algorithms available to implement various types of fused lasso estimation methods.
	For instance, Tibshirani et al. \cite{tibshirani2005sparsity} utilized a two-stage dynamic algorithm that transforms variables into a form involving differences between positive and negative parts.
	However, this algorithm appears to be inefficient.
	Li et al. \cite{li2014linearized} utilized a proximal alternating direction method of multipliers (ADMM) \cite{Glowinski1975SurLP,1976A}, which involves linearizing the quadratic terms within the augmented Lagrangian function to enable closed-form solutions.
	Li et al. \cite{li2018efficiently} employed a level-set method to effectively address the fused lasso problems. Their approach involved solving a sequence of regularized least squares subproblems and applying the semismooth Newton method.
	On the implementation of the estimation method (\ref{admod}), Xin et al. \cite{xin2022robust} utilized some auxiliary variables to transform  the model into a separable equality constrained problem, facilitating efficient resolution via ADMM. 
	However, despite ADMM's usual numerical efficiency, it requires solving a high-dimensional linear system within its framework.

	To overcome this challenge, we shift our focus to the dual formulation of \eqref{admod}, which is structured with two separable non-smooth function blocks. 
	Compared to the algorithm of Xin et al. \cite{xin2022robust}, the dual approach notably decreases the dimensionality of auxiliary variables, resulting in a reduction in computational complexity.
	We employ the semi-proximal ADMM (spADMM) algorithm to minimize the dual model. This method dynamically generates proximity terms to reformulate subproblems into proximal mappings, enabling closed-form solutions for each subproblem.
	Moreover, it should be noted that, from the excellent work of Fazel et al. \cite{fazel2013hankel}, the convergence of the spADMM  we employ is easily guaranteed. 
	Finally, we also conduct extensive numerical experiments which demonstrates that the spADMM exhibits outstanding performance.
	
	The remainder of the paper is organized as follows: In Section \ref{sec2}, we provide a brief review of basic concepts and preliminary knowledge in optimization and statistics. In Section \ref{sec3}, we focus on designing a semi-proximal ADMM to solve problem \eqref{admod} from a dual perspective. In Section \ref{sec4}, we present the results of numerical experiments and provide some comparisons to demonstrate the performance of the proposed algorithm. Finally, we conclude the paper with some remarks in Section \ref{sec5}.

	\section{Preliminaries}\label{sec2}
	In this section, we  quickly recall  some basic concepts  and  preliminary results in convex optimization. 
	Let $\mathbb{R}$ be a finite-dimensional real Euclidean space endowed with an inner product $\langle \cdot, \cdot \rangle$ and its induced norm $\|\cdot\|$.
	Let $f:\mathbb{R}\rightarrow (-\infty, +\infty]$ be a closed proper convex function, the Fenchel conjugate function of $f$, denoted as $f^{*}$, is
	\begin{align*}
		f^{*}(y): =\sup_x\{\langle y, x \rangle-f(x) \ | \ x\in \mathbb{R} \}=-\inf_x\{f(x)-\langle y, x \rangle \ | \ x\in \mathbb{R}\}.
	\end{align*}
	For a nonempty closed convex set $\mathcal{C}$, the symbol $\delta_{\mathcal{C}}(x)$ represents an indicator function over $\mathcal{C}$ such that $\delta_{\mathcal{C}}(x)=0$ if $x\in\mathcal{C}$ and $+\infty$ otherwise.
	The conjugate of an indicator function $\delta_{\mathcal{C}}(x)$ is named support function  defined by
	$\delta^{*}_{\mathcal{C}}(x)=\sup \{ \langle x, y\rangle | y\in \mathcal{C}\}$.
	It is not hard to deduce that the Fenchel conjugate of $\|x\|_{p}$ is $\|x\|_{p}^*=\delta_{\mathcal{B}^{(1)}_{q}}(x)$ where $\mathcal{B}^{(1)}_{q}:=\{x | \|x\|_q\leq 1\}$ and $1/p+1/q=1$.
	
	The Moreau-Yosida regularization  \cite{moreau1962fonctions} of a closed proper convex function $f$ at $x\in \mathbb{R}$ is defined as
	\begin{equation}\label{MY}
		\psi_{f}(x):=\min_{y\in \mathbb{R}} \{f(y)+\frac{1}{2}\| y-x\|^{2}\}.
	\end{equation}
	For any $x\in \mathbb{R}$, problem (\ref{MY}) has a unique optimal solution, which is called the proximal point of $x$ associated
	with $f$, or the proximal mapping operator,  i.e.,
	\begin{equation*}
		\mbox{prox}_{f}(x):=\mbox{arg} \min_{y\in \mathbb{R}} \{f(y)+\frac{1}{2}\| y-x\|^{2}\}.
	\end{equation*}
	In particular, the proximal point of $x$ associated with an indicator function $\delta_{\mathcal{C}}(x)$ reduces to the metric projection of $x$ onto $\mathcal{C}$, i.e.,
	\begin{equation}\label{proxf}
		\text{prox}_{\delta_{\mathcal{C}}}(x)=\text{arg}\min_{y\in \mathbb{R}} \{\delta_{\mathcal{C}}(y)+\frac{1}{2}\| y-x \|^{2} \} = \text{arg}\min_{y\in \mathcal{C}} \{\frac{1}{2}\| y-x \|^{2}\}=\Pi_{\mathcal{C}}(x).
	\end{equation}
	
	The relationship between the proximal point associated with a function and its conjugate is described as Moreau's identity \cite[Theorem 35.1]{Rock35}
	\begin{equation}\label{moreauid}
		\prox_{tf}(x)+t\prox_{f^{*}/t}(x/t)=x.
	\end{equation}
	The proximal mapping for some norm functions can be computed flexibly utilizing  the  Moreau's identity,  which is essential in the subsequent analysis. For example:
	\begin{itemize}
		\item[(i)] Let $f(x)=\mu\|x\|_1$with $\mu>0$, then $f^{*}(x^*)=\delta_{\B^{(\mu)}_{\infty}}(x^*)$ with ${\B^{(\mu)}_{\infty}}(x^*):=\{x^* \ | \ \|x^*\|_{\infty}\leq \mu\}$ and
		$$
		\prox_f(x^*)=x^*-\Pi_{\B^{(\mu)}_{\infty}}(x^*) \quad \text{with} \quad
		(\Pi_{B^{(\mu)}_{\infty}}(x^*))_i=\left\{\begin{array}{ll}
			x^*_i, & \text { if }  |x^*_i|\leq\mu,\\
			\operatorname{sign}(x^*_i) \mu, & \text { if }  |x^*_i|>\mu.
		\end{array}\right.
		$$
		\item[(ii)]\cite{Dolan2001BenchmarkingOS,yu2013decomposing} Let $f(x)=\mu g^*(x)$, where $g^*$ is the conjugate function of $g$ in the form of $g(\beta)=\lambda_1\| \beta\|_1 + \lambda_2\sum_{j=2}^{p}|\beta_{j} - \beta_{j-1}|$, then
		\begin{equation}\label{proxp}
			\prox_{\mu g^*}(x)= x-\mu\prox_{g/\mu}(x/\mu)  \quad \text{with} \quad \prox_g(x)=\prox_{\lambda_1\|\cdot\|_1}(x_{\lambda_2}(y)),
		\end{equation}
		where $x_{\lambda_2}(y)$ is the proximal mapping of $\lambda_2\sum_{j=2}^{p}|\beta_{j} - \beta_{j-1}|$. 
	\end{itemize}

	\section{Dual Model and SPADMM}\label{sec3}
	It should be noted that there are many efficient numerical approaches that can be employed to solve the fused lasso penalized least squares model as mentioned previously. However,  numerical algorithms for the Huber loss in the form of (\ref{admod}) have received limited investigation. 
	Therefore,  it is important to develop an efficient and robust algorithm to implement the estimation method (\ref{admod}). 
	In this section, we aim to derive the dual formulation of the model given in \eqref{admod} and design an effective and robust algorithm to solve it.
	
	Define a multivariate Huber function by empirical loss
	$$
	\mathcal{H}_{\tau}(x)=\frac{1}{n}\sum_{i=1}^{n}h_{\tau}(x_i),
	$$
	where $h_{\tau}(\cdot)$ is defined in \eqref{huber}. Obviously, the multivariate Huber function is convex according the convexity of $h_{\tau}(\cdot)$. 
	For convenience, we simplify the term $\sum_{j=2}^{p}|\beta_{j} - \beta_{j-1}|$ as $\| D  \beta\|_1$, where
	$$
	D=
	\begin{bmatrix}
		-1  &   1    & 0       & \cdots\ & 0\\
		\vdots  & \vdots & \ddots  & \ddots  & \vdots\\
		0   &  0    & \cdots\ & 1       &0 \\
		0   &   0    & \cdots\ & -1      & 1 \\
	\end{bmatrix}
	$$
	is named a difference operator matrix in $\mathbb R^{(p-1)\times p}$. 
	Using the difference operator matrix,  the model (\ref{admod}) enjoys an ideal separable structure by using some auxiliary variables  and hence can be solved efficiently by employing ADMM. 
	It should be noted that the ADMM implemented by Xin et al. \cite{xin2022robust} requires solving a  $p\times p$ linear system. 
	However, it is evident that the computational cost of solving this linear system will significantly increase with the dimension $p$, thereby having an adverse effect on the algorithm’s efficiency.
	Hence, it is very necessary to improve the performance of the ADMM used by Xin et al. \cite{xin2022robust} using some technical skills. 
	
	Let $ z:=  y -  X \beta$,	we rewrite (\ref{admod}) equivalently as the following model 
	\begin{equation}\label{pmod}
		\begin{array}{rl}
			\min\limits_{ z\in \mathbb{R}^n, \beta\in \mathbb{R}^p} & 
			\mathcal{H}_{\tau}( z) + p(\beta),\\ 
			\text{s.t.}  &   z=  y -  X \beta,
		\end{array}
	\end{equation}
	where $p(\beta):=\lambda_1\| \beta\|_{1} + \lambda_2\| D  \beta\|_1$. 
	The Lagrangian function associated with  \eqref{pmod} is
	$${\cal L}( z, \beta;{u}):= \mathcal{H}_{\tau}( z) + p(\beta)-\langle  u,  z-  y +  X \beta \rangle,$$
	where $ u\in  \mathbb{R}^n$ is a  multiplier. The Lagrange dual function of \eqref{pmod} is to minimize  ${\cal L}( z, \beta;{u})$ with  respect to $z$ and $\beta$, that is 
	\begin{eqnarray*}
		D(u):=\inf_{z,\beta} {\cal L}( z, \beta;{u})=-\mathcal{H}^*_{\tau}( u) - p^*({X}^{\top} u)+\langle  u, y\rangle,
	\end{eqnarray*}
	where $\mathcal{H}^{*}_{\tau}(\cdot)$ and $p^{*}(\cdot)$ represent the conjugate functions of $\mathcal{H}_{\tau}(\cdot)$ and $p(\cdot)$, respectively.   The Lagrangian dual problem of the original \eqref{pmod} is to maximize the dual function $D(u)$, which can equivalently be written as the following minimization problem:
	\begin{equation}\label{du0}
		\min\limits_{ u\in \mathbb{R}^n} 
		\mathcal{H}^*_{\tau}( u) +p^*({X}^{\top} u)-\langle  u, y\rangle.
	\end{equation}
	Here, it is a trivial task to calculate the form of $\mathcal{H}_\tau^{*}(\cdot)$ that $\mathcal{H}_\tau^{*}(u)=\frac{n}{2}\| u\|^2 + {\delta_{\mathbb B^{\tau/n}_{\infty}}}(u)$.
	
	Let $ v: ={X}^{\top} u$, then \eqref{du0} reformulates as the following separable form:
	\begin{equation}\label{dmod1}
		\begin{array}{rl}
			\min\limits_{ u\in \mathbb{R}^n, v\in \mathbb{R}^p} & 
			\mathcal{H}^*_{\tau}( u) + 	p^*( v)-\langle  u, y\rangle,\\ 
			\text{s.t.}  &  v= {X}^{\top} u.
		\end{array}
	\end{equation}
	Clearly, the model \eqref{dmod1} has separable structure in terms of both the objective function and  the constraint, and thus, it falls into the framework of ADMM. 
	The key idea of ADMM  is minimizing augmented Lagrangian function $\mathcal{L}_{\sigma}( u,  v;  w)$ regarding to $u$ and $v$ individually, and then update its multiplier with the latest values $u$ and $v$. 
	The augmented Lagrangian function associated with dual problem \eqref{dmod1} takes the following form
	\begin{equation}
		\label{lag}
		\mathcal{L}_{\sigma}( u,  v;  w):=\mathcal{H}^*_{\tau}( u) + 	p^*(v)-\langle u, y\rangle-\langle w, v-{X}^{\top} u\rangle+\frac{\sigma}{2}\|v-{X}^{\top} u\|^2,
	\end{equation}
	where $w\in \mathbb R^p$ is multiplier associated with the constraint. 
	We note that, due to the constant term ${X}$, the $u$-subproblem in ADMM does not have a closed-form solution. 
	To address this issue, we add proximal terms to the $u$- and $v$-subproblems. Consequently, with given $(u^k, v^k;w^k)$, the next iteration $(u^{k+1}, v^{k+1};w^{k+1})$ is generated through the following iterative scheme:
	
	\begin{equation}\label{admm}
		\left\{
		\begin{array}{l}
			u^{k+1}= \argmin_{ u}\mathcal{L}_{\sigma}( u,  v^k;  {w}^k)+\frac{1}{2}\| u-  u^k\|^2_{\cal S},\\[3mm]
			v^{k+1}= \argmin_{ v}\mathcal{L}_{\sigma}( u^{k+1},  v;  {w}^k)+\frac{1}{2}\| v-  v^k\|^2_{\cal T},\\[3mm]
			{w}^{k+1}={w}^k-\sigma\rho( v^{k+1}-{X}^{\top} u^{k+1}),
		\end{array}
		\right.
	\end{equation}
	where $\rho \in (0, (1+\sqrt{5})/2)$ is a step length, $\cal S$ and $\cal T$ are semi-definite matrices.

	We note that two subproblems are present in (\ref{admm}), each of which constitutes the primary computational burden. 
	In the following, we demonstrate that analytic solutions are permitted, thereby facilitating an easily implementable framework. 
	Let $I_n$ be the identity matrix.
	Firstly, choose $\S:= (\eta-n) I_{n}-\sigma X X^{\top}$ where $\eta>0$ is a positive scalar such that $\S$ is semi-positive definite. 
	Fix $u:=u^k$ and $v:=v^k$, then we get for every $k=0, 1, \ldots$, that
	\begin{eqnarray*}
		u^{k+1} &:=& \argmin_{ u}\mathcal{L}_{\sigma}( u,  v^k;  {w}^k)\\
		&=& \argmin_{ u}\frac{n}{2}\| u\|^2 + {\delta_{\mathbb B^{\tau/n}_{\infty}}}(u)- \langle u, y\rangle + \langle w^k, X^{\top} u\rangle + \frac{\sigma}{2}\| v^k- X^{\top} u\|^2 + \frac{1}{2}\| u-  u^k\|^2_{\cal S}\\
		& = & \Pi_{\mathbb{B}_{\infty}^1}^{\tau/n}(\eta^{-1}(y- Xw^k+\sigma X v^k+{\cal S} u^k))
	\end{eqnarray*}
	where the last equality is from the definition of proximal mapping in \eqref{proxf}.
	Secondly, choose $\mathcal{T}:= 0$, $u:=u^{k+1}$, and $w:=w^k$, we get 
	for every $k=0, 1,\ldots,$ that
	\begin{eqnarray*}
		v^{k+1}&:=& \argmin_{ v}\mathcal{L}_{\sigma}( u^{k+1},  v;  {w}^k).\\
		&=& \argmin_v p^*( v)-\langle w^k, v\rangle +\frac{\sigma}{2}\| v- X^{\top} u^{k+1}\|^2\\
		&=& \prox_{p^*}^{1/\sigma}( X^{\top} u^{k+1}+\frac{w^k}{\sigma})\\
		&=& X^{\top} u^{k+1} + \sigma^{-1} w^k - \sigma^{-1} \prox_{\sigma p}(\sigma X^{\top} u^{k+1} + w^k)
	\end{eqnarray*}
	where the last equality is also from the definition of proximal mapping and Moreau's identity in \eqref{moreauid}.

	From \eqref{proxp}, it is evident that computing $\prox_{p^*}(x)$ relies on efficiently calculating $x_{\lambda_2}(\cdot)$. 
	We note that there are a lot of algorithms have been reviewed in the recent paper \cite{pustelnik2017proximity}.
	In our subsequent numerical experiments, we employ the algorithm proposed by Condat \cite{condat2013direct}, which has undergone extensive testing across various environments. In summary, we are ready to state the full steps of the semi-proximal ADMM (named spADMM)  while it is used to solve the dual problem (\ref{dmod1}).
	\begin{framed}
		\noindent
		{\bf Algorithm $1$: spADMM}
		\vskip 1.0mm \hrule \vskip 1mm
		\noindent
		\begin{itemize}
			\item[Step 0.] Choose a starting point $( u^0, v^0;{w}^0)\in \mathbb{R}^n\times \mathbb{R}^{p}\times \mathbb{R}^{p}$. Choose a positive constant $\eta>0$ such that $\mathcal{S}$ be positive semi-definite.
			Choose positive constants $\sigma>0$, $\lambda>0$, and $\rho\in(0,(1+\sqrt{5})/2)$. For $k=0,1,\ldots$, do the following operations iteratively.
			\item[Step 1.] Given $u^k$, $v^k$, and $w^k$, compute 
			$u^{k+1}:=\Pi_{\mathbb{B}_{\infty}^{\tau/n}}(\eta^{-1}(y- Xw^k+\sigma X v^k+{\cal S} u^k))$.
			\item[Step 2.] Given $u^{k+1}$, $v^k$, and $w^k$, compute
			$$
			v^{k+1}:= X^{\top}u^{k+1}+w^k/\sigma-\sigma^{-1}\text{Pro}_{\sigma p}(\sigma X^{\top}u^{k+1}+w^k).
			$$
			\item[Step 3.] Given $u^{k+1}$, $v^{k+1}$, and $w^k$, compute
			${w}^{k+1}:={w}^k-\sigma\rho( v^{k+1}-{X}^{\top} u^{k+1})$.
		\end{itemize}
	\end{framed}
	
	For clarity in understanding the above algorithm, we give a couple of remarks.

	\begin{remark}
		In our later experiments, we terminate the iterative process of spADMM the relative error of two continue points is sufficiently small, i.e., $\mbox{RelErr}< \mbox{Tol}$, where RelErr is defined as 
		\begin{equation}\label{stop}
			\mbox{RelErr} :=\frac{\|\beta^{k+1}-\beta^k\|_2}{1+\|\beta^k\|_2}.
		\end{equation}
	\end{remark}

	\begin{remark}
		The statistical theory of the estimation method  \eqref{admod} has been studied by Xin et al. \cite{xin2022robust}, who establishes the nonasymptotic convergence rates of the estimation method in a high-dimensional setting.  Under some  bounded moment and restricted eigenvalue conditions, they established an upper bound for the difference between the estimated $\hat{\beta}$ and the ground truth $\beta^*$: for any $t>0$, it holds that 
		$$
		\|\hat{ \beta} - \beta^*\|_2 \lesssim C_n\sqrt{\frac{s}{(n/t)^{\min\{2\delta/(1+\delta), 1\}}}},
		$$  where $C_n>0$ is a constant, $s$ represents the cardinality of the  true support of $\beta^*$, and $\delta$  is associated with the bounded moment condition for random error, defined as $ \max\limits_{1\leq i\leq n}\mathbb{E}(\mid\varepsilon_i\mid^{1+\delta}) < \infty$.
	\end{remark}

	To conclude this subsection, we present the convergence result of spADMM without proof. For the details on its proof, one may refer to \cite[Theorem B.1]{fazel2013hankel}.
	\begin{theorem}(\cite[Theorem B.1]{fazel2013hankel})\label{theo1}
		Suppose that the sequence $\{(u^{k}, v^{k}; w^{k})\}$ is generated by spADMM from
		an initial point $(w^0, v^0; w^0)$.
		If $\rho\in(0, (1+\sqrt{5})/2)$ and $\eta>0$ is chosen such that $\mathcal{S}$ being positive semi-definite.
		Then the sequence $\{(u^{k}, v^{k})\}$ converges to an optimal
		solution of  (\ref{dmod1}) and $\{w^{k}\}$ converges to an optimal solution of the corresponding primal problem \eqref{pmod}.
	\end{theorem}

	\section{Numerical experiment} \label{sec4}
	In this section, we  test the effectiveness and accuracy of the proposed estimation method on sparse linear regression with  Huber loss and fused lasso penalty using typical synthetic data and  real biology  data. 
	All the experiments are conducted on a PC with Microsoft Windows 11 and MATLAB R2022b, featuring an Intel Core i7 CPU at 1.80 GHz and 16 GB of memory. 
	It should be noted that, Xin et al. \cite{xin2022robust} employed a symmetric Gauss-Seidel ADMM (named FHADMM) to evaluate the robustness and effectiveness of the estimation method. They also made comparisons with fused lasso penalized least squares estimation method. 
	Therefore, in this test, we only focus on evaluating the numerical performance of spADMM, with a particular emphasis on comparing it to FHADMM.
	
	\subsection{Simulated example}
	At the first place, we describe the data generation process used in this test.  
	The response variables obey the form $y=X\beta^* +\epsilon$ but with different distribution of random noise:
	\begin{itemize}
		\item[(i)] Gaussian noise: $\epsilon\sim\mathcal{N}(0,0.05^2)$;
		\item[(ii)] $t$ distribution noise with degrees of freedom $1.5$:  $\epsilon\sim  t_{1.5}$;
		\item [(iii)] Lognormal noise:  $\epsilon\sim\log\mathcal{N}(0, 2^2)$;
		\item[(iv)]Laplace noise: $\epsilon\sim\text{Laplace}(0, 1)$.
	\end{itemize}
	We note that these types of noise can be  categorized into two groups: the light-tailed noise (i), and the heavy-tailed noise(ii, iii, and iv). 
	In addition, the noise (iv) is allowed to test the sensitivity of the estimated method on  outliers.
	The rows of $X$ are given by $n$ independent Gaussian vector $\mathcal N(0,\Sigma )$, where $\Sigma_{i,j}=0.5^{|i-j|}$  for $1\leq i,j\leq p$. The vector of true coefficient $\beta^*$ is generated by  
	$$ 
	\beta_i^*=\begin{cases}
		1,& \lceil\frac{p}{5}\rceil+1\leq i\leq \lceil\frac{2p}{5}\rceil,\\[1mm]
		-1.5,&\lceil\frac{3p}{5}\rceil+1\leq i\leq \lceil\frac{3p}{4}\rceil,\\[1mm]
		0, &\text{otherwise},
	\end{cases}
	$$
	where ``$\lceil\cdot\rceil$'' denoted as a ceiling function. Obviously, the $\beta^*$ takes the following form 
	$$
	\beta^*=(0,\dots,0,1,\dots,1,0,\dots,0,-1.5,\dots,-1.5,0,\dots,0).
	$$ 
	To comprehensively evaluate the numerical performance of the estimation method, we use the following three measurements:
	\begin{description}
		\item[(1)] $ \mbox{RLNE}:=\frac{\|\hat{\beta}-\beta^*\|_2}{\|\beta^*\|_2}$, where $\hat{\beta}$ is an estimated sparse coefficient and RLNE describes the accuracy of estimation.
		\item[(2)] $\mbox{MAE}:=\frac{1}{n}\sum\limits_{i=1}^{n}|y_i-X_i^{\top}\hat{\beta}|$, in which, MAE describes the accuracy of predictions.
		\item[(3)] $\mbox{MSE}:=\frac{1}{n}\sum\limits_{i=1}^{n}(y_i-X_i^{\top}\hat{\beta})^2$, in which,
		MSE also measures the accuracy of predictions, but it is more sensitive to outliers.
	\end{description}
	These measurements allow us to capture the algorithm's performance from different aspects and then provide a more thorough assessment among different numerical algorithms.
	In the following numerical experiments, we stop the iterative process of each algorithm if RelErr is sufficiently small, or the maximum iteration number is achieved.

	\subsubsection{Behavior of spADMM}
\begin{figure}
		\centering
		\subfloat[$\lambda_1$, NT=1]{\includegraphics[width=4cm]{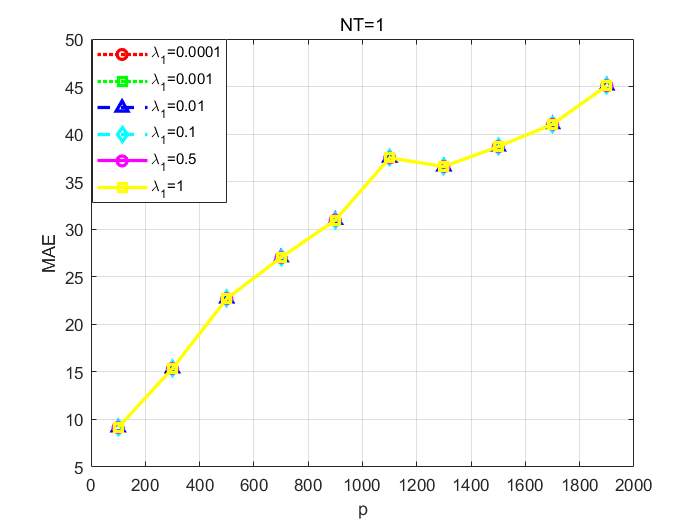}}
		\subfloat[$\lambda_1$, NT=2]{\includegraphics[width=4cm]{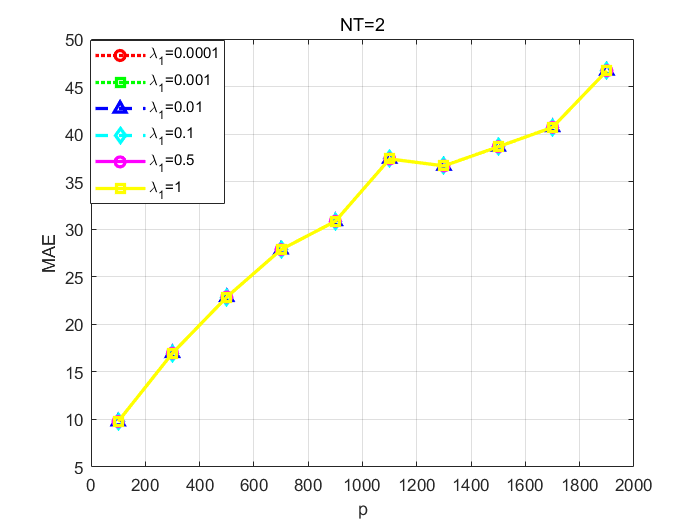}}
		\subfloat[$\lambda_1$, NT=3]{\includegraphics[width=4cm]{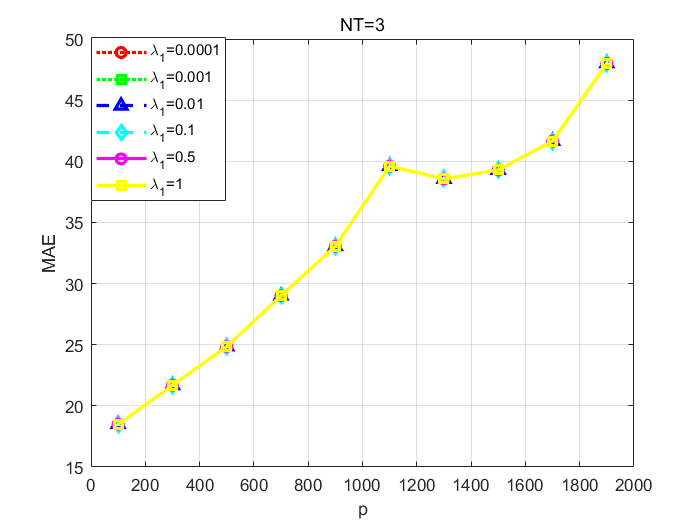}}
		\subfloat[$\lambda_1$, NT=4]{\includegraphics[width=4cm]{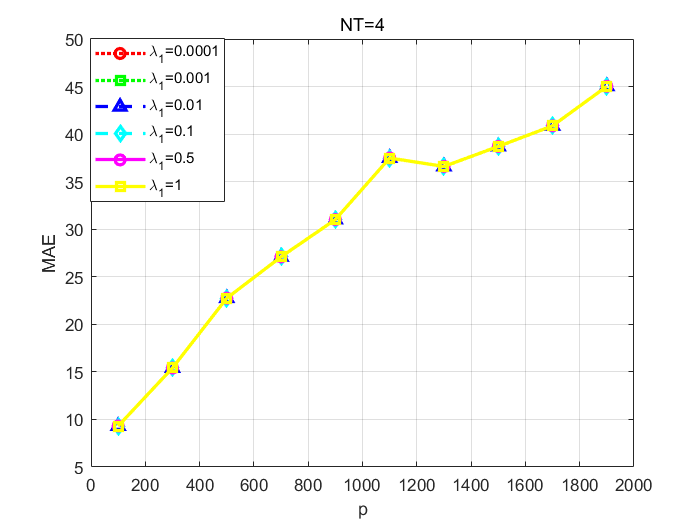}}
		\\
		\subfloat[$\lambda_2$, NT=1]{\includegraphics[width=4cm]{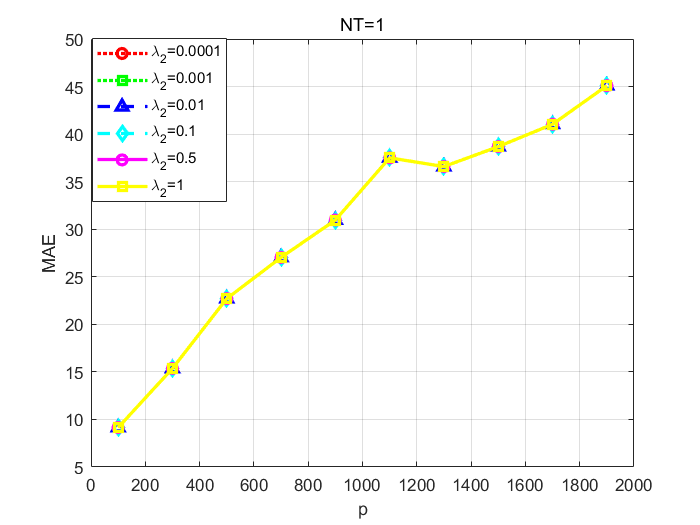}}
		\subfloat[$\lambda_2$, NT=2]{\includegraphics[width=4cm]{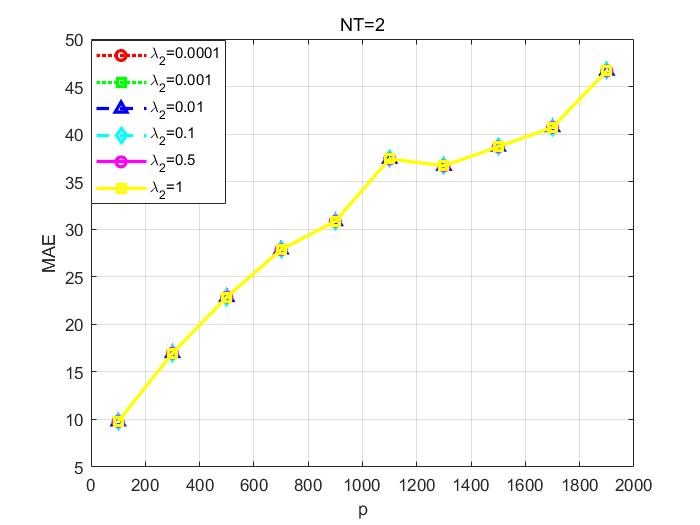}}
		\subfloat[$\lambda_2$, NT=3]{\includegraphics[width=4cm]{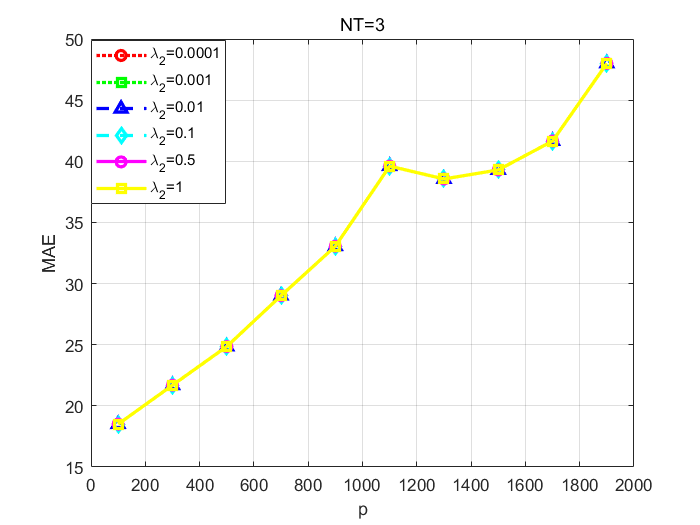}}
		\subfloat[$\lambda_2$, NT=4]{\includegraphics[width=4cm]{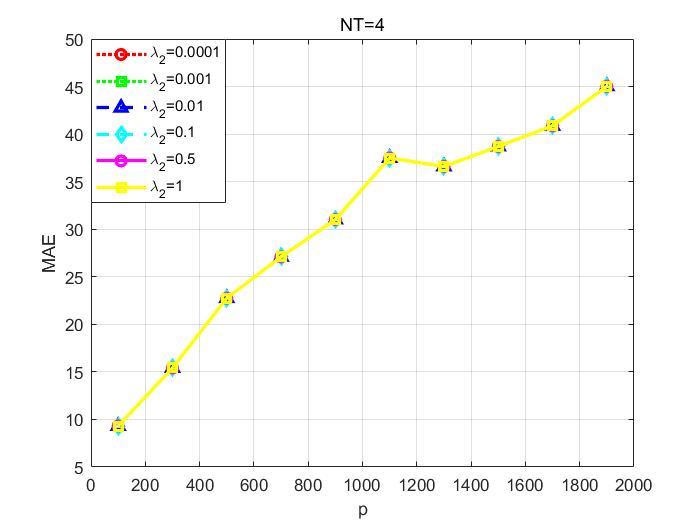}}
		\\
		\subfloat[$\tau$, NT=1]{\includegraphics[width=4cm]{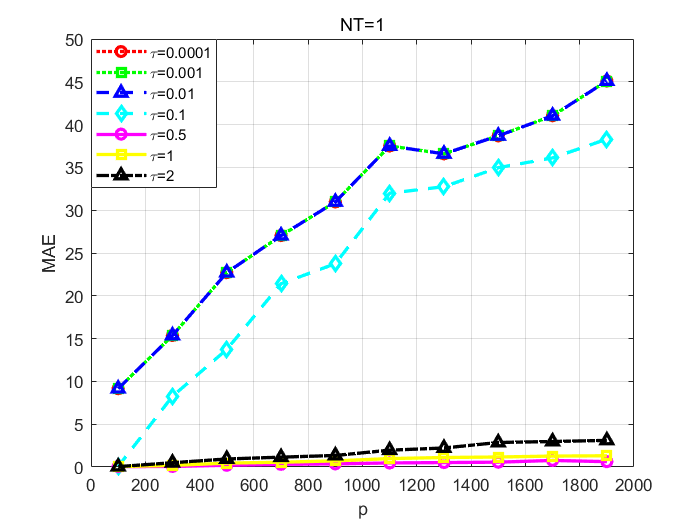}}
		\subfloat[$\tau$, NT=2]{\includegraphics[width=4cm]{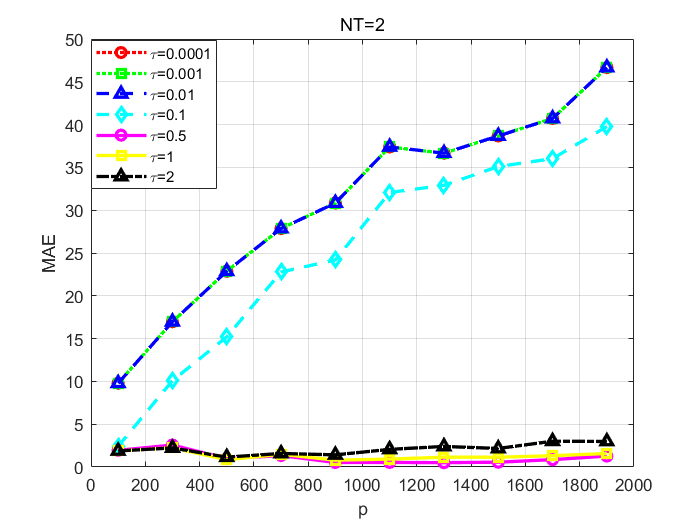}}
		\subfloat[$\tau$, NT=3]{\includegraphics[width=4cm]{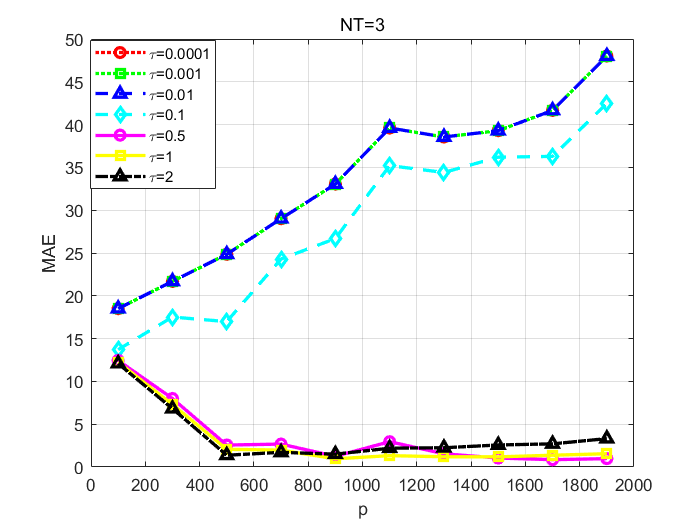}}
		\subfloat[$\tau$, NT=4]{\includegraphics[width=4cm]{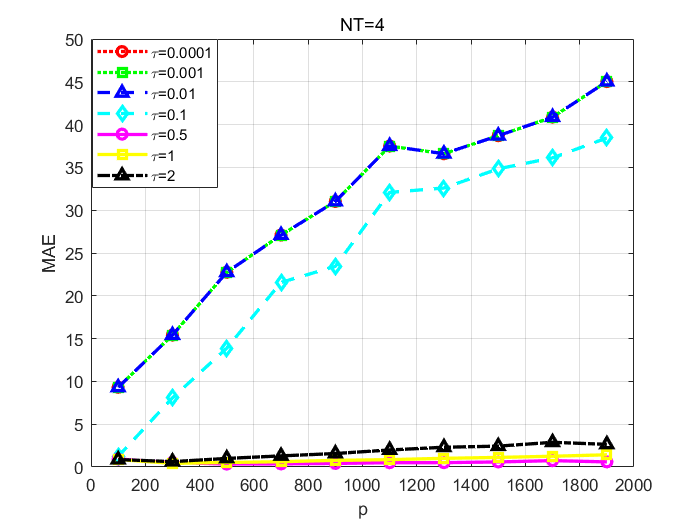}}
		\caption{The performance of Algorithm 1 under different parameters with various dimensions and noise types, where ``NT" represents the type of noise with values ranging from 1 to 4 corresponding to the specific noise introduced earlier in (i) to (iv).}
		\label{fig:F1}
	\end{figure}

	To showcase the numerical performance of spADMM, we conducted various tests to evaluate its estimation effectiveness from different perspectives. 
	Firstly, we investigate the sensitivity of proposed approach to model's parameters: the robustification parameter $\tau$ as well as the regularization parameters $\lambda_1$ and $\lambda_2$, under different noise types mentioned above. 
	In this test, the algorithm is terminated when the maximum iteration number reached $1000$. 
	The sample size is set to $n=200$, and sample's dimensions increase from $100$ to $2000$. 
	Here, we employ MAE to evaluate  algorithm's performance and  plot the relationship between tuning parameters and MAE in Figure \ref{fig:F1}. 
	
	From Figure \ref{fig:F1}, we can clearly see that spADMM is insensitive to the values of $\lambda_1 $ and $ \lambda_2 $. 
	In general, regardless of the specific values of $ \lambda_1 $ and $ \lambda_2 $, the MAE values roughly exhibit an upward trend as the dimension $p$ increases across the four types of noise. 
	However, this phenomenon is particularly evident in the panel related to $\tau$, especially concerning lognormal noise. Specifically, the MAE values are notably high when the magnitude of $ \tau$ is less than $0.1$, indicating a significant impact of $ \tau$ on algorithm performance. Therefore, in the later experiments, we keep $\lambda_1$ and $\lambda_2$ fixed while focusing on tuning $ \tau$.

\begin{figure}
		\centering
		\subfloat[$200\times 500$, NT=1]{\includegraphics[width=4cm]{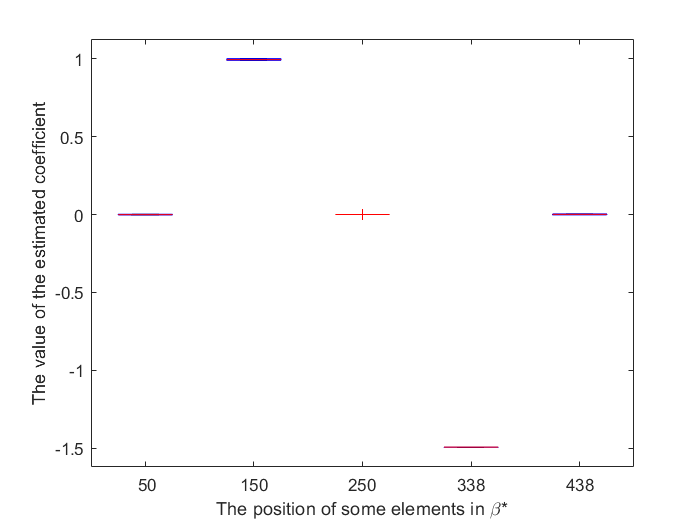}}
		\subfloat[$200\times 500$, NT=2]{\includegraphics[width=4cm]{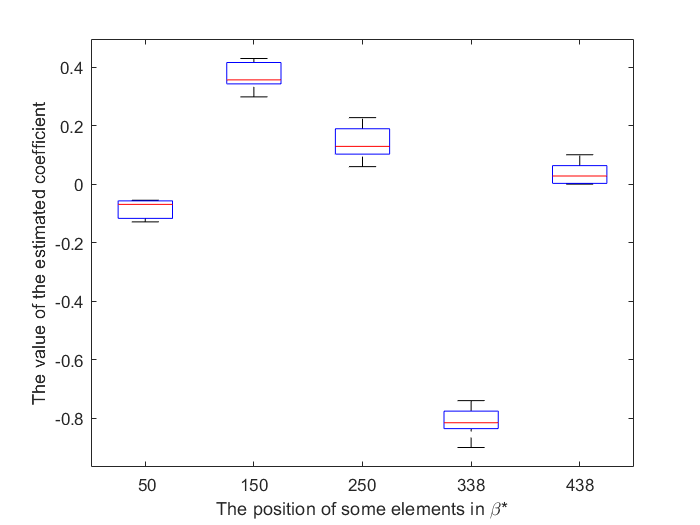}}
		\subfloat[$200\times 500$, NT=3]{\includegraphics[width=4cm]{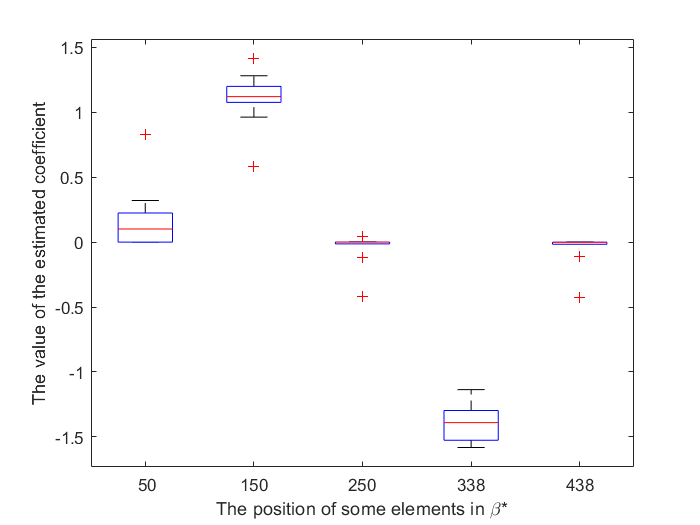}}
		\subfloat[$200\times 500$, NT=4]{\includegraphics[width=4cm]{Figure/2005003.png}}
		\\
		
		\subfloat[$500\times 800$, NT=1]{\includegraphics[width=4cm]{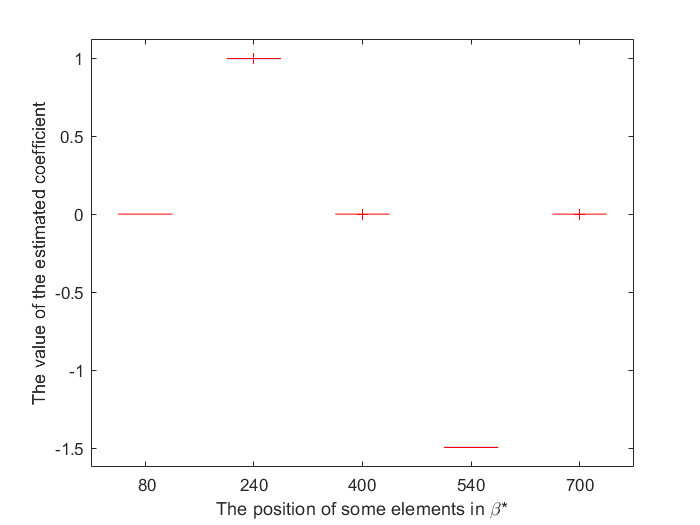}}
		\subfloat[$500\times 800$, NT=2]{\includegraphics[width=4cm]{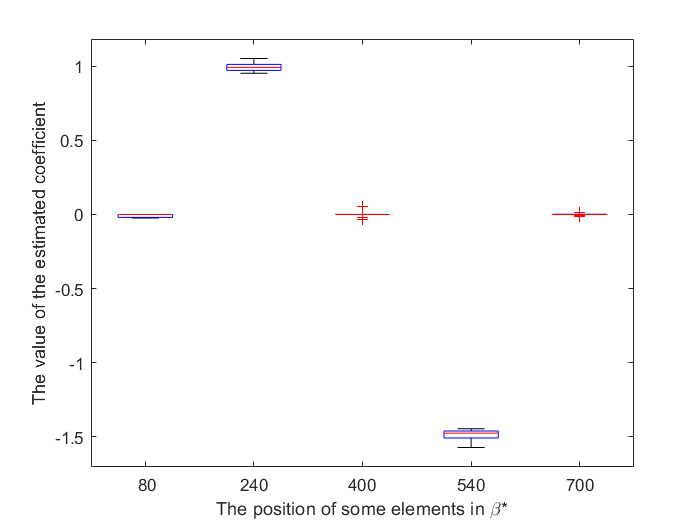}}
		\subfloat[$500\times 800$, NT=3]{\includegraphics[width=4cm]{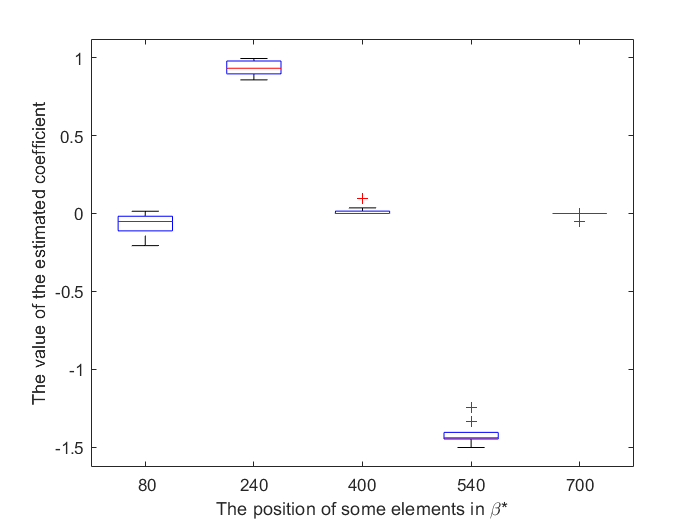}}
		\subfloat[$500\times 800$, NT=4]{\includegraphics[width=4cm]{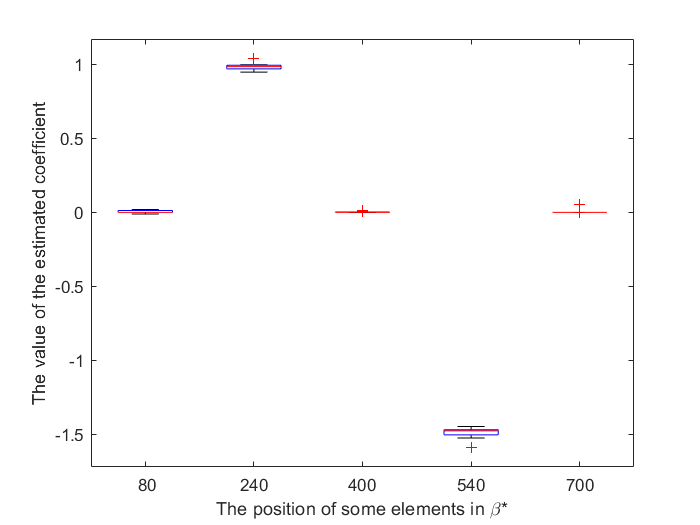}}
		\\
		\subfloat[$800\times 1100$, NT=1]{\includegraphics[width=4cm]{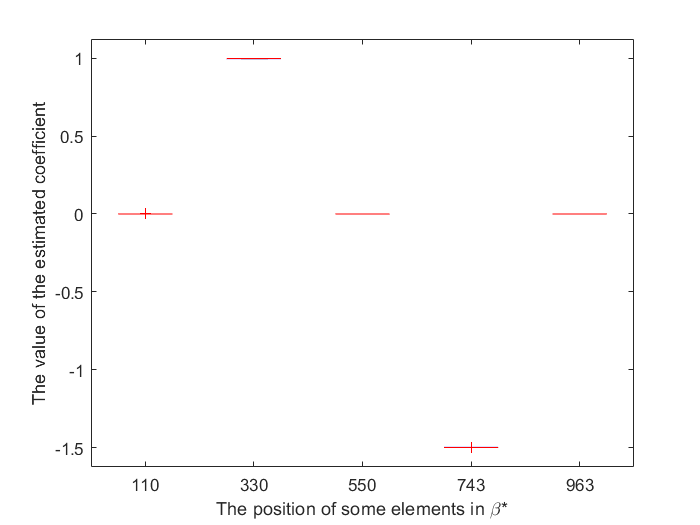}}
		\subfloat[$800\times 1100$, NT=2]{\includegraphics[width=4cm]{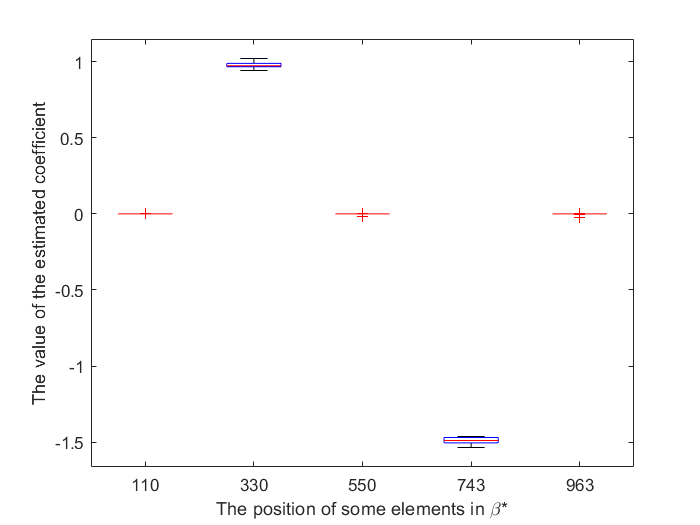}}
		\subfloat[$800\times 1100$, NT=3]{\includegraphics[width=4cm]{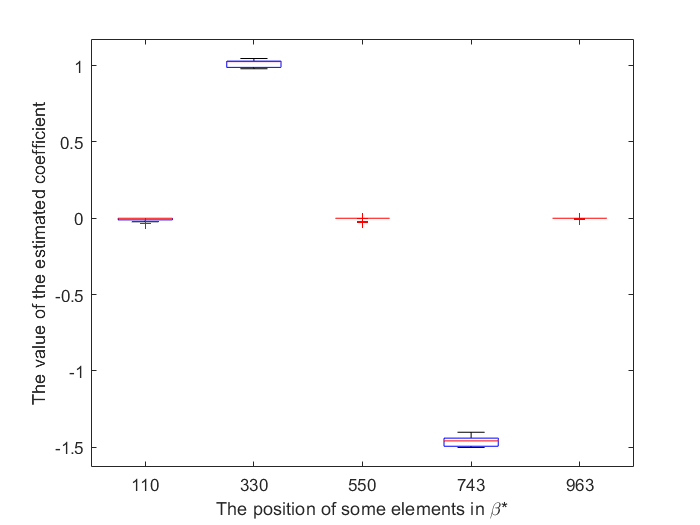}}
		\subfloat[$800\times 1100$, NT=4]{\includegraphics[width=4cm]{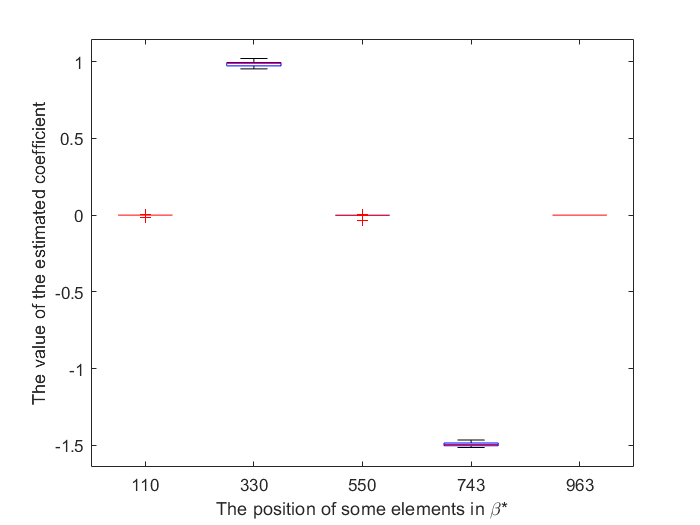}}
		\caption{Estimation performance of Algorithm 1 under different noise and sample size.}	\label{fig:F2}
	\end{figure}

	Now, we focus on assessing the algorithm's stability and estimation accuracy under small and large sample sizes. Specifically, the sample size are set to $200\times 500$, $500\times 800$, and $800\times 1100$, respectively. We choose $\lambda_1 = \lambda_2 = 0.01$, $\tau=0.5$,  and terminate algorithm's process if iteration number is achieved  $1000$. We list the box plots in Figures \ref{fig:F2} which illustrates the estimation performance of the spADMM at  specific positions,  based on  $10$ independent experiments. 
	Here, the position we focus on are $i=\lceil\frac{p}{10}\rceil$, $\lceil\frac{3p}{10}\rceil$, $\lceil\frac{p}{2}\rceil$,  $\lceil\frac{27p}{40}\rceil$, and $\lceil\frac{7p}{8}\rceil$. From Figure \ref{fig:F2}, it is evident that the spADMM accurately identifies the  position of zero and non-zero elements across all independent experiments.  
	Despite  the box plots (panels(b)--(d)) show that experimental results  are somewhat dispersed when  the size of design matrix $X$ is  $200\times 500$,  their quantile points remain within an error margin of  $\pm 0.5$ from the true values. 
	Overall, these observations demonstrate  that spADMM is able to successfully achieve the task of parameters estimation and variables selection.

	\subsubsection{Performances comparisons}	
	Given the demonstrated robustness and predictability of the algorithm FHADMM for the \eqref{admod}, we  specifically compare its numerical performance with  spADMM. These comparisons contain the computational accuracy and efficiency of both algorithms using both simulated and real data.
	
	In the simulation test, we construct the predictor matrices  $X$ with different size. The parameter $\tau$ is  selected through grid search from the set $\{0.001,0.01,0.1,0.5,1, a\sqrt{n/\text{log}(p)}\}$, where $a$ is incremented from $0.4$ to $1.5$ with an interval of $0.05$. The regularization parameters remain consistent with those employed in earlier experiments. 
	In addition to using the maximum number of iterations $5000$, we also use $\mbox{RelErr}<1.0e-3$ as a termination criterion. 
	We compare spADMM and FHADMM from six aspects: number of iterations (Iter), CPU time (Time) in seconds, RelErr, RLNE, MAE, and MSE. 
	The numerical results of both algorithms are reported in Table \ref{tab:simu}.
	
\begin{table}[htbp]
	\centering {\tiny\caption{Numerical performance of spADMM(a) and FHADMM(b) on  simulated data.}
		\begin{tabular}{ccccccc}
			\hline
			\multirow{2}{*}{\textbf{Size}} & \textbf{Iter} &\textbf{Time} & \textbf{RelErr} & \textbf{RLNE} & \textbf{MAE} & \textbf{MSE} \\
			&$ a\mid b$&$ a\mid b $&$ a\mid b $&$ a\mid b$&$ a\mid b $&$ a\mid b$\\
			\hline
			\multicolumn{7}{c}{Gaussian noise}\\
			$200\times 500 $&$689\mid 3182$&$0.591\mid 1.273$&$9.81e-04\mid 9.89e-04$&$7.66e-01\mid 7.21e-01$&$2.37e-01\mid 2.45e-01$&$2.17e-02\mid 2.23e-02$\\
			$500\times 800 $&$327\mid 1689$&$1.267\mid 4.359$&$9.97e-04\mid 9.70e-04$&$8.49e-01\mid 3.79e-01$&$9.25e-02\mid 1.45e-01$&$5.89e-03\mid 9.82e-03$\\
			$800\times 1100 $&$344\mid 574$&$2.321\mid 3.202$&$9.84e-04\mid 9.28e-04$&$7.68e-03\mid 8.52e-02$&$2.17e-01\mid 1.89e-01$&$9.68e-03\mid 9.26e-03$\\
			\multicolumn{7}{c}{t-distribution noise}\\
			$200\times 500$&$ 812\mid  3572 $&$0.494\mid  1.626  $&$9.95e-04\mid 9.98e-04$&$7.20e-01\mid 7.28e-01$&$2.36e-01\mid 2.45e-01$&$2.19e-02\mid 2.22e-02$\\	
			$500\times 800 $&$408\mid 652$&$1.522\mid 1.544$&$9.87e-04\mid 9.96e-04$&$9.08e-02\mid 8.40e-02$&$7.83e-03\mid 8.84e-03$&$2.14e-03\mid 2.19e-03$\\
			$800\times 1100 $&$371\mid 603$&$2.111\mid 3.354$&$9.85e-04\mid 9.77e-04$&$6.66e-02\mid 7.10e-02$&$7.69e-03\mid 8.29e-03$&$1.37e-03\mid 1.39e-03$\\
			\multicolumn{7}{c}{Lognormal noise}\\
			$200\times 500 $&$679\mid 3740$&$0.423\mid 1.780$&$1.00e-03\mid 9.98e-04$&$7.78e-01\mid 8.28e-01$&$2.09e-01\mid 2.37e-01$&$1.93e-02\mid 2.15e-02$\\
			$500\times 800 $&$394\mid 876$&$1.442\mid 2.218$&$9.95e-04\mid 1.00e-03$&$2.90e-01\mid 9.64e-02$&$9.77e-03\mid 1.25e-02$&$2.49e-03\mid 2.56e-03$\\
			$800\times 1100 $&$403\mid 2355$&$8.435\mid 22.889$&$9.98e-04\mid 8.82e-04$&$9.21e-01\mid 5.82e-01$&$3.64e-03\mid 7.81e-03$&$1.34e-03\mid 1.37e-03$\\ 
			\multicolumn{7}{c}{Laplace noise}\\
			$200\times 500 $&$714\mid 2362$&$3.100\mid 4.269$&$9.96e-04\mid 9.95e-04$&$7.62e-01\mid 7.96e-01$&$2.34e-01\mid 2.37e-01$&$2.18e-02\mid 2.12e-02$\\
			$500\times 800 $&$96\mid 730$&$0.987\mid 4.388$&$7.87e-04\mid 9.74e-04$&$9.83e-01\mid 9.96e-01$&$2.56e-01\mid 2.47e-01$&$1.44e-02\mid 1.38e-02$\\
			$800\times 1100 $&$394\mid 796$&$8.525\mid 8.464$&$9.98e-04\mid 9.76e-04$&$7.49e-01\mid 6.46e-02$&$1.20e-01\mid 2.23e-01$&$5.77e-03\mid 1.01e-02$\\
			\hline
		\end{tabular}
		\label{tab:simu}
	}
\end{table}
	
	From the results presented in table \ref{tab:simu}, it is observed that both algorithms meet the termination condition within the maximum iteration step. 
	Furthermore, we see that the low error values measured by RLNE, MAE, and MSE demonstrate the satisfactory performance of both algorithms in estimation and prediction, irrespective of the type of noise. 
	This phenomenon also indicates that the estimation method \eqref{admod} is capable of dealing with the   data contaminated by  heavy-tailed noise.
	
	Notably, compared to FHADMM, spADMM consistently requires significantly fewer iterations and less computing time in  high-dimensional scenario.
	But for the other four measurements characterizing the estimation quality and  prediction accuracy, the corresponding value of both algorithms have its own potential advantages. In most cases, spADMM is more competitive with the majority of instances for MAE and MSE. 
	These observations highlight the spADMM's advantages in time efficiency in the high-dimensional scenarios.

	\subsection{Real data examples}
	In this part, we further test the effectiveness of our estimation method using three datasets from biomedical research, and then compare the performance of spADMM and FHADMM. The first real-world dataset is leukemia data\footnote{\url{https://hastie.su.domains/CASI\_files/DATA/leukemia.html}} \cite{golub1999molecular} which consists of  $7129$ genes (dimension) and $72$ samples. 
	In these samples, $47$ patients were diagnosed with acute lymphocytic leukemia (ALL), while the remaining $25$ samples were diagnosed with acute myelogenous leukemia (AML). 
	The second real-world dataset is liver cancer\footnote{\url{https://www.cancer.gov/about-nci/organization/ccg/research/structural-genomics/tcga}}  \cite{wheeler2017comprehensive} which consists of $19,255$ genes and $116$ samples with an equal number of patients and healthy individuals.
	The third real-world dataset is bladder cancer data\footnote{\url{https://www.cancer.gov/about-nci/organization/ccg/research/structural-genomics/tcga}} \cite{cancer2014comprehensive}  which comprises $19,211$ genes and $40$ samples, and in these samples, there are $21$ patients diagnosed with bladder cancer.
	
	In this test, we roughly screen out genes with significant differences after performing hierarchical clustering, and then run spADMM and FHADMM using these filtered datasets. 
	We set the maximum iterations number as $1,500$ and choose `Tol' as $1.0e-3$, $1.0e-4$, and $1.0e-5$, respectively. 
	The numerical results and the size of filtered datasets are all reported in Table \ref{tab:true}.
	It is noteworthy that the response variable for these datasets is binary, hence we choose `Accuracy' and `Recall rate' as measurement to make the more accurate evaluations. 
	Here, the `Accuracy' presents the ratio of the correct number of samples predicted by a  classifier to the total number of samples, and the `Recall rate' denotes the proportion of true positive samples correctly identified  out of the total actual positive samples \cite{Tharwat2020ClassificationAM}.
	\begin{table}[htbp]
	\centering {\scriptsize\caption{Numerical results of spADMM(a) and FHADMM(b) on real data.}
	\resizebox{18cm}{!}{
		\begin{tabular}{ccccccccc}
			\toprule[1pt]
			\multirow{2}{*}{\textbf{Data name}}&	\multirow{2}{*}{\textbf{Tol}}& \textbf{Iter}& \textbf{Time}& \textbf{RelEr}r&\textbf{MAE}&\textbf{MSE}&\textbf{Accuracy}&\textbf{Recall}  \\
			& &$a\mid b$& $a\mid b$& $a\mid b$& $a\mid b$& $a\mid b$& $a\mid b$& $a\mid b$\\
			\midrule
			\multirow{3}{*}{$\text{leukemia} \atop (72,3707)$}
			&1.00e-03&$11\mid 28$&$0.040\mid 3.786$&$9.98e-04\mid 7.90e-04$&$7.70e-01\mid 6.96e-03$&$8.22e-01\mid 7.55e-05$&$0.722\mid 1.000$&$0.872\mid 1.000 $\\
			&1.00e-04&$408\mid 33$&$1.946\mid 3.837$&$1.00e-04\mid 7.36e-05$&$3.93e-01\mid 7.02e-03$&$2.17e-01\mid 7.70e-05$&$0.972\mid 1.000$&$1.000\mid 1.000 $\\
			&1.00e-05&$9162\mid 5486$&$40.042\mid 202.530$&$1.00e-05\mid 1.00e-05$&$2.16e-01\mid 1.34e-01$&$7.03e-02\mid 2.88e-02$&$1.000\mid 1.000$&$1.000\mid 1.000$\\\midrule
			\multirow{3}{*}{$\text{liver} \atop (116,2596)$}
			&1.00e-03&$13\mid 190$&$0.027\mid 4.200$&$8.11e-04\mid 8.54e-04$&$9.34e-01\mid 9.22e-01$&$9.06e-01\mid 8.97e-01$&$0.759\mid 0.500$&$0.914\mid 1.000 $\\
			&1.00e-04&$32\mid 230$&$0.063\mid 4.987$&$3.62e-05\mid 9.01e-05$&$9.34e-01\mid 9.35e-01$&$9.06e-01\mid 9.10e-01$&$0.724\mid 0.828$&$0.914\mid 0.776 $\\
			&1.00e-05&$36\mid 1240$&$0.070\mid 20.617$&$8.97e-06\mid 9.94e-06$&$9.34e-01\mid 9.34e-01$&$9.06e-01\mid 9.08e-01$&$0.724\mid 0.828$&$0.914\mid 0.775$\\
			\midrule
			\multirow{3}{*}{$\text{bladder} \atop (40,2541)$}
			&1.00e-03&$50\mid 30$&$0.043\mid 1.329$&$9.96e-04\mid 7.33e-04$&$6.97e-01\mid 1.79e-01$&$5.77e-01\mid 6.38e-02$&$0.900\mid 1.000$&$1.000\mid 1.000 $\\
			&1.00e-04&$1198\mid 752$&$0.978\mid 13.267$&$1.00e-04\mid 8.95e-05$&$5.47e-01\mid 2.48e-01$&$3.77e-01\mid 1.18e-01$&$1.000\mid 1.000$&$1.000\mid 1.000 $\\
			&1.00e-05&$12003\mid 4617$&$9.899\mid 69.989$&$1.00e-05\mid 1.00e-05$&$5.46e-01\mid 2.17e-01$&$3.76e-01\mid 1.06e-01$&$1.000\mid 1.000$&$1.000\mid 1.000 $\\
			\bottomrule[1pt]
		\end{tabular}}
		\label{tab:true}
	}
\end{table}

	From the computational results presented in this table, we see that  spADMM and FHADMM meet the accuracy requirements within $1,500$ iteration steps, and produce comparable values of `Accuracy' and `Recall'. 
	Additionally, the MAE value of spADMM is slightly lower than that of FHADMM, and this difference is more evident when testing on the leukemia dataset.
	From the column of `Time', we see that spADMM is generally faster than FHADMM in the case of getting high degree of accuracy measured by the termination criteria.  
	Here, it's worth noting that these datasets have dimensions that exceed their sample sizes. From this analysis, we can conclude that in high-dimensional contexts, spADMM delivers similar robustness and accuracy as FHADMM, while achieving better computational efficiency.
	
	\section{Conclusion}\label{sec5}
	In this paper, we considered a robust estimation method and  proposed an efficient optimization algorithm for high-dimensional sparse linear regression problems so as to address  the challenges posed by heavy-tailed errors and outliers. 
	In this estimation method, we considered Huber regression function for robust estimation and used the fused lasso penalty to encourage the sparsity of coefficients and their successive differences. 
	To avoid solving a large linear equation appear in the algorithmic framework of Xin et al. \cite{xin2022robust}, we designed a computational skill and proposed a semi-proximal ADMM based on dual formulation.
	We conducted a series of numerical experiments using both simulated and real data, and compared the performance with FHADMM.
	These numerical results showed that the proposed estimation method is  robust and that  the implementation algorithm is generally computationally efficient. 
	\section*{Acknowledges}
	The work of Y. Xiao is supported by the National Natural Science Foundation of China (Grants No. 12471307 and 12271217).

	\bibliographystyle{ieeetr}
	\bibliography{huberref}
\end{document}